# The modification of Einstein`s gravitational field equation following from the energy conservation law

**Roald Sosnovskiy**
Technical University, 194021, St. Petersburg, Russia
E-mail:rosov2@yandex.ru

**Abstract.** The cause of an infringement in GR of a gravitational field energy conservation law is investigated . The equation of a gravitational field not contradicting to the energy conservation law is suggested. This equation satisfy to the Einstein's requirement of equivalence of all energy kinds as sources of a gravitational field. This equation is solved in paper for cosmic objects. It is showed, that results for some objects - for black holes and gravitating strings -essentialy differ from such for Einstein's equation, have the symple meaning and do not contradictions.

**Introduction.** The paper is devoted to the problem of $t_i^k$ gravitational field energy. This problem takes special place in the relativity theory as on its importance as on difficulty. Einstein's "energy - moment pseudotensor " $t_i^k$ cannot be interpreted as the quantity adequately describing the energy - moment of a gravitational field. So, for the elementary case, cylindrically symmetric static field, $t_0^0 = 0$, that is erroneous. In this case a pseudotensor Landau-Lifshits has also incorrect value [2], [5]. Many authors offered different variants pseudotensor [7], [9],[11], however all of them appeared unsatisfactory.

In the review devoted to a problem of energy of a gravitational field [8], are in details considered and systematized attempts to find a satisfactory variant pseudotensor $t_i^k$. It was made the conclusion that it is impossible to find such pseudotensor describing density of the energy - moment. There is a generally accepted opinion, that the energy - moment of a gravitational field should be local, and that the local law of energy-moment conservation must be fulfilled [8].

In paper [8] are considered other approaches to the decision of a problem of the gravitational field energy - moment also. Attempts to refuse of pseudotensors use were made by introduction of additional connections, use of higher derivatives or a 4 - formalism, development of alternative theories of gravitation [8], [10], [12]. However, these attempts were unsuccessful.

There is the opinion [3] that to descript an energy quantity it is difficult because, for it cannot to be measured. It is possible to agree with this opinion, since these make difficulties in searching of alternative approaches. However, in [5] there was suggested the mental experiment, allowing find the gravitational field energy density in little volume. Used calculations based on the energy conservation law. This results used in presented paper in order to solve the gravitational field energy conservations problem for static systems. Such problem could be solved in principle for systems having Killing`s vectors [4] including static systems. In the presented paper made attempt to use these results in order to solve a problem of conservation of energy – moment, at least for static systems.

In a case of the relative movement of material bodies, in all space there can be streams of energy. Localization of energy and the description of fields in such systems demand using of more complicated approaches and in the present paper is not considered. The consideration is limited to static systems.

In this paper are studied problems:

-the analysis of the cause of a gravitational field energy conservation law infringement

-the modification of Einstein`s gravitational field equation

-the solution of the modified equation for spherically and cylindrically symmetrical field

-the solution of the system of a gravitational field equation and a field masse density equation for cosmical objects.

**2. The cause of the infringement of a gravitational field energy conservation law.** The Einstein´s initial assumption in the time of an elaboration of the gravitational field equation was that Ricci tensor is equal 0 for free from material bodies part of space

$$R_{\mu\nu} = 0 \tag{1}$$

This equation is obvious for the total empty space. However, it is not obvious for the free from matter part of a space, containing the matter in others parts. Then this assumption is equivalent to the assumption that gravitational field energy not generate gravitational field. However, such assumption contradict to requirement, that "the gravitational field energy must be equivalent in full to all others energy forms" [1].

From (1) follow Einstein`s equation [2]

$$R_i^k - \frac{1}{2}R = \frac{8\pi G}{c^4} T_i^k \tag{2}$$

Here $T_k^i$ is an energy-moment density tensor of matter. The equation (2) is asymmetrical with regard to matter and field energy. However, the pseudotensor $t_i^k$ it follows from so equation. Thus, just the assumption about the gravitational field energy not generate such field is the cause of a breach of a gravitational field energy conservation law.

**3. Modification of gravitation field equation.** In order to remove the cause of a breach a gravitational field energy conservation law it is suggested attach (2) to form

$$G_i^k = R_i^k - \frac{1}{2}R = \frac{8\pi G}{c^4}(T_i^k + \hat{t}_i^k) \tag{3}$$

Here $\hat{t}_i^k$ is a tensor (*not a pseudotensor*) of the gravitational field energy-moment density and $\hat{t}_0^0$ is a gravitational field energy density from [5]. The tensor $\hat{t}_i^k$ it is not possible to found from (3). The gravitational field energy density $\hat{t}_0^0$ it is possible calculate from equation [5]

$$\frac{\partial}{\partial x^1}\left(\frac{\partial M_f(x^1, M_0)}{\partial M_0} + \frac{1}{2}\ln g_{00}\right) = 0 \tag{4}$$

Here $x^1$ is the coordinate coinciding with trajectory of test particles falling with infinitesimal velocity (that is by test particles motion energy dispersion);

$dM_f(x^1,M_0)$ is the gravitational field mass inside tube of trajectories in a layer $dx^1$;

$dM_0$ is the increase of matter masse on the body surface inside a this tube;

$g_{ik}$ – is the metric tensor.

It is obvious that

$$\hat{t}_0^0 = c^2 \mu_f(x) \tag{5}$$

Here $\mu_f(x)$ is the field mass density.

Must be also fulfilled conservation laws

$$\hat{t}_{i;k}^k = \frac{1}{\sqrt{-g}} \frac{\partial}{\partial x^k} \left( \sqrt{-g} \hat{t}_i^k \right) - \frac{1}{2} g_{rs,i} \hat{t}^{rs} = 0 \tag{6}$$

and

$$\frac{\partial \sqrt{-g}\left(T_i^k + \hat{t}_i^k\right)}{\partial x^k} = 0 \tag{7}$$

For each point of a field it is necessary to count $g_{ik}, M_f$, tensor $\hat{t}_i^k$. As will be shouwn further, the equations (3)-(7) completely definie $\hat{t}_i^k$ for static systems. Tensor $\hat{t}_i^k$ for moved systems is not calculated in this paper.. But, it is necessery to note, that its behaviour in such systems is normally. For observer's coordinate system moving relatively of field source with constant speed v : $M_{fv} = \dfrac{M_f}{\sqrt{1 - v^2/c^2}}$ and consequently $\hat{t}_{0v}^0 = \dfrac{\hat{t}_0^0}{\sqrt{1 - v^2/c^2}}$. For free falling observer's coordinate system, the gravitational field is absent and $M_{fa} = 0, \hat{t}_{0a}^0 = 0$.

For material bodies system with a variable configuration cannot use directly of the equations (3), (4). Such systems not consider therefore.

**4. The gravitational field energy -moment tensor for the spherically symmetrical systems.** For spherical symmetry the interval $ds^2$ hat a form

$$ds^2 = g_{ii}\left(dx^i\right)^2 \tag{8}$$

where

$$dx^0 = dt,\ dx^1 = dr,\ dx^2 = d\theta,\ dx^3 = d\varphi \tag{9}$$
$$g_{22} = r^2,\ g_{33} = -\sin^2\theta\, r^2 \tag{10}$$

Equation (3) for vacuum with a gravitational field hat then the form

$$G_i^i = R_i^i - \frac{1}{2} R = \frac{8\pi G}{c^4} \hat{t}_i^i \tag{11}$$

The solutions of (11) is [13]

$$g_{11} = -\left(1 - \frac{8\pi G}{c^4 r} \int_0^r r^2 \hat{t}_0^0 dr\right)^{-1} \tag{12}$$

$$g_{00} = -\frac{1}{g_{11}} \exp\left\{\frac{8\pi G}{c^4} \int_0^r r g_{11}\left(\hat{t}_1^1 - \hat{t}_0^0\right) dr\right\} \tag{13}$$

The field mass density equal

$$\mu(r) = \frac{M_{f,1}(r)}{4\pi r^2} \tag{14}$$

where $M_f(r)$ is the field mass of a spherical area with radius r. If $r_0$ is the radius of a material body and $M_0$ is his mass, $r_0 > \dfrac{2GM_0}{c^2}$, then

$$\int_0^r r^2 \hat{t}_0^0 dr = \frac{c^2}{4\pi}[M_0 + M_f(r)] \tag{15}$$

and

$$g_{11} = -\left[1 - \frac{2GM(r)}{c^2 r}\right]^{-1}, \quad M(r) = M_0 + M_f(r) \tag{16}$$

From the conservation requirements (6), (7) follow

$$\hat{t}_1^1 = 0 \tag{17}$$

and from (13)

$$g_{00} = \left(1 - \frac{2GM(r)}{c^2 r}\right) \exp\left\{\frac{2G}{c^2} \int_{r_0}^r \frac{M_{f,1} dr}{r\left(1 - \frac{2GM(r)}{c^2 r}\right)}\right\} \tag{18}$$

Then

$$\frac{g_{00,1}}{g_{00}} = \frac{2GM(r)}{c^2 r^2}\left[1 - \frac{2GM(r)}{c^2 r}\right]^{-1} \tag{19}$$

From (6) one can to obtain

$$\hat{t}_2^2 = \hat{t}_3^3 = -\frac{r g_{00,1}}{4 g_{00}} \hat{t}_0^0 \tag{20}$$

**5. The gravitational field energy-moment tensor for the cylindrically symmetrical systems.** For $ds^2$ in (8) where

$$dx^0 = dt, \ dx^1 = dr, \ dx^2 = d\varphi, \ dx^3 = dz \tag{21}$$

and by the assumption

$$g_{11} = -1, \ g = -r^2, \ \frac{g_{ii,1}}{g_{ii}} = \frac{a_i}{r}, \ g_{ii} = \left(\frac{r}{r_0}\right)^{a_i} \tag{22}$$

one can to obtain

$$a_1 = 0, \sum_i a_i = 2, \sum_i a_{i,1} = 0 \tag{23}$$

The Riccy covariant tensor [2]

$$R_{ii} = \Gamma^l_{ii,l} - \Gamma^l_{il,i} + \Gamma^l_{ii}\Gamma^m_{lm} - \Gamma^m_{il}\Gamma^l_{im} \tag{24}$$

From (22) and (24) the mixed Riccy tensor is equal

$$R_i^i = \frac{1}{2r} a_{i,1} + \frac{\delta_i^1}{r^2}\left(\frac{1}{4}\sum_i a_i^2 - 1\right) \tag{25}$$

Then

$$R = \sum_i R_i^i = \frac{1}{r^2}\left(\frac{1}{4}\sum_i a_i^2 - 1\right) = R_1^1 \tag{26}$$

For $i = 1$ the tensor component $\hat{t}_1^1$ in consequence of conservation law (6) is equal

$$\hat{t}_1^1 = 0 \tag{27}$$

and then from (3)

$$G_1^1 = R_1^1 - \frac{1}{2} R = 0 \tag{28}$$

In consequence of (25), (26), (27)

$$\sum_i a_i^2 = 4 \tag{29}$$

If a gravitational field mass density µ(r) is known then is known a tensor component $\hat{t}_0^0$ and from (11), (25), (26), (29)

$$G_0^0 = \frac{1}{2r} a_{0,1} = \frac{8\pi G}{c^2} \mu(r) \tag{30}$$

$$\hat{t}_0^0 = \frac{M_{zf,r} c^2}{2\pi r} \tag{31}$$

If the matter cylinder radius is $r_0$ and his linear mass density is $M_{z0}$ then an equation (30) give

$$a_0(r) = a_0(r_0) + \frac{8G}{c^2}[M_z(r) - M_{z0}] \tag{32}$$

$$M_z(r) = M_{zf}(r) + M_{z0} \tag{33}$$

Here $M_{zf}(r)$ is the field mass of a cylindrical area with radius r.

From [5] on the material cylinder surface

$$\frac{a_0(r_0)}{r_0} = \frac{g_{00,1}(r_0)}{g_{00}(r_0)} = \frac{4GM_{z0}}{c^2 r_0} \tag{34}$$

Then (31) give

$$\frac{g_{00,1}}{g_{00}} = \frac{a_0(r)}{r} = \frac{4G}{c^2 r}(M_{z0} + 2M_{zf}) \tag{35}$$

The tensor's $\hat{t}_i^i$ components $\hat{t}_2^2, \hat{t}_3^3$ one can to obtain from the conservation law (6)

$$\hat{t}_2^2 = \frac{c^4 a_{2,1}}{16\pi Gr}; \hat{t}_3^3 = \frac{c^4 a_{3,1}}{16\pi Gr} \tag{36}$$

Here $a_2$, $a_3$ are equal from (23),(29) to

$$a_2 = \frac{2-a_0}{2}\left[1 + \sqrt{1 + \frac{4a_0}{2-a_0}}\right] \tag{37}$$

$$a_3 = \frac{2-a_0}{2}\left[1 - \sqrt{1 + \frac{4a_0}{2-a_0}}\right] \tag{38}$$

**6. The solution of equations (3), (4) system for spherically symmetrical obects.** Gravitatin mass of a field inside of the sphere equal: $M(r) = M_0 + M_f(r, M_0)$. This mass creates a field outside of sphere. It is possible to write down the equation (4) in the form of

$$\frac{dM}{dr} + \frac{1}{2}\int_0^{M_0} \frac{g_{00,r}}{g_{00}} d\tilde{M}_0 = 0 \tag{39}$$

Here $\tilde{M}_0 = 0 \div M_0$ and $g_{00} = g_{00}(r, \tilde{M}_0)$.

From (39) follows

$$\frac{dM_f}{dr} = \frac{M_0}{2r} + \frac{c^2}{2G}\ln\left(\frac{1 - \frac{2G}{c^2}\frac{(M_f + M_0)}{r}}{1 - \frac{2G}{c^2}\frac{M_r}{r}}\right) \tag{40}$$

For objects with spherical symmetry as argument x and functions depending on it quantity are used

$$x = \ln\left(\frac{r}{r_0}\right); \quad y = \frac{2GM}{c^2 r}; \quad u = \frac{M}{M_0}; \quad pu = \frac{1}{M_0}\frac{dM}{dx} \tag{41}$$

For quantities in (41) this equation hat a form

$$dy = \left[ -y + \frac{1}{2} y_0 e^{-x} - \frac{1}{2} \ln\left(1 + \frac{y_0 e^{-x}}{1-y}\right) \right] dx \qquad (42)$$

Boundary condition x=0, y=y$_0$. Relative value of mass of system on distance r from the center of system equal

$$u = \frac{M}{M_0} = \frac{ry}{r_0 y_0} \qquad (43)$$

If $y_0 \ll 1$ that, representing a logarithm in (42) in the polynomial form, it is possible to lead this equation to a form

$$\frac{dM_f}{d(1/r)} - \frac{2G}{c^2} M_0 M_f \approx \frac{2G}{4c^2} M_0^2 \qquad (44)$$

The solution of this equation

$$u = \frac{M}{M_0} \approx \frac{1}{4}\left\{ 3 + \exp\left[ y_0 \left(1 - \frac{r_0}{r}\right)\right] \right\} \qquad (45)$$

For Sun $M_0 = 1{,}99 \cdot 10^{33}$ g, $r_0 = 6{,}96 \cdot 10^{10}$ and $y_0 = 4{,}24 \cdot 10^{-6} \ll 1$.

From the equation (45) follows, that at greater distances r » r$_0$ (order of space horizon)

$$\frac{M}{M_0} \approx 0{,}99999$$

For the white dwarf Sirius B $M_0 = 1{,}97 \cdot 10^{33}$ g, $r_0 = 5{,}42 \cdot 10^8$ and $y_0 = 5{,}45 \cdot 10^{-6} \ll 1$. From (45) follows, that at such distances

$$\frac{M}{M_0} \approx 0{,}999986$$

Thus, for these stars influence of field mass gravitation is extremely insignificant.

For black holes $y_0 = 1$ and the formula (42) is used. Computer calculation is made with step dy = 0,001. As in the formula (42) relative quantities x, y, y$_0$ contain only, results of this calculation are suitable for black holes of any sizes. On the Fig.1.a dependence $y = 2GM/c^2 r$ and $u = M/M_0$ from $x = \ln(r/r_0)$ is presented. One can see, at great values x M/M$_0$ = 0,71, i.e. influence of gravitation of a field on full mass of system is essential.

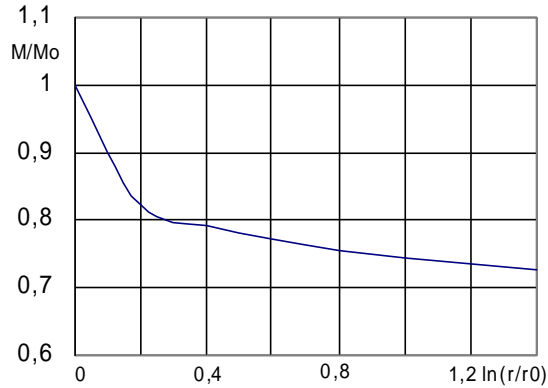

Fig.1. Relative mass u = M/M$_0$ as a function of ln (r/r$_0$) for black holes

The density of a field energy from (14) can be presented in the form of

$$\frac{\hat{t}_0^0}{\overline{T}_0^0} = \frac{1}{3}\exp(-3x) \cdot \frac{du}{dx} \tag{46}$$

Here $\overline{T}_0^0 = \frac{3M_0 c^2}{4\pi r_0^3}$ - average density of energy in a spherical source of a field.

On Fig.2 dependence pu=du/dx and $t = \frac{\hat{t}_0^0}{\overline{T}_0^0}$ from x is presented

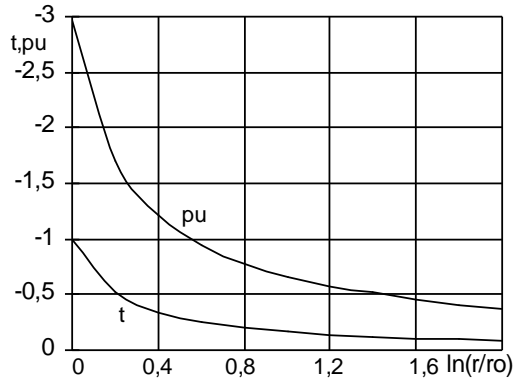

Fig.2. Relativ mass density $t = \hat{t}_0^0 / \overline{T}_0^0$ and $pu = dM / M_0 dx$ as a function of x=ln(r/ro) for black holes.

From the diagram Fig.2 in a range ln (r/r$_0$) = 0 ÷ 0,2 quantities $\hat{t}_0^0$ and du/dx = dM/M$_0$dx change approximately in 2 times. It essentially changes the metrics of space near to a black hole.

The received results, apparently, are suitable for massive black object in the center of a galaxy which is considered in [16], [17] as a black hole.

Components $g_{00}$, $g_{11}$ of metrical tensor can be calculated from (16), (17) by M=M$_0$·u.

**7. The solution of equations (3), (4) system for cylindrically symmetrical obects.** For cylindrically symmetric systems and at boundary conditions $r = r_0, \frac{dM_{zf}}{dM_{z0}} = 0$ this equation have a form

$$\frac{dM_{zf}}{dM_{z0}} + \frac{1}{2}\ln g_{00}(r, M_{zf}, M_{z0}) = \frac{1}{2}g_{00}(r_0, M_{z0}) \tag{47}$$

The solution of this equation

$$M_{zf} = -\frac{1}{2}M_{z0} + \frac{c^2}{8G\ln\frac{r}{r_0}}\left[1 - \exp\left(-4\frac{GM_{z0}}{c^2}\ln\frac{r}{r_0}\right)\right] \tag{48}$$

For objects with cylindrical symmetry quantities are used

$$x_c = b \cdot \ln(\frac{r}{r_0}); b = \frac{4GM_{z0}}{c^2}; u_c = \frac{M_z}{M_{z0}} \tag{49}$$

From the equation (48) using (49) it is possible to receive

$$u_c = \frac{1}{2}\left(1 + \frac{1-e^{-x_c}}{x_c}\right) \tag{50}$$

$$\frac{du_c}{dx_c} = \frac{-1 + (1 + x_c)e^{-x_c}}{2x_c^2} \tag{51}$$

Relative mass $u_c = M_z/M_{z0}$ as a function of $x_c = \ln(r/r_0)$ is presented on Fig.3.

On can see, that a influence of gravitation field mass on a full mass is essentially.

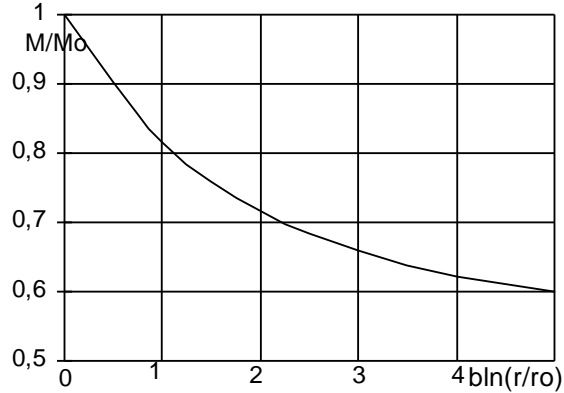

.3. Relative mass $M_z/M_{zo}$ as a function of $x_c = b\cdot\ln(r/r_0)$ for cylindrically symmetrical system.

The relation of a field energy density to a energy density of the body is equal from (31) and (51):

$$\frac{\hat{t}_0^0}{\overline{T}_{c0}^0} = \frac{1}{2}\exp(-\frac{2x}{b})\frac{du_c}{dx_c} \tag{52}$$

Here $\overline{T}_{co}^0 = \frac{3M_{z0}c^2}{4\pi r_0^3}$ is an average density of energy in a gravitating body. On the diagram Fig.4 dependence of quantity $\frac{\hat{t}_{c0}^0}{\overline{T}_0^0}$ from $\frac{r}{r_0}$ for value $M_{z0}=1022$ g/cm [18] i.e. $b = 3\cdot 10^{-6}$ is resulted.

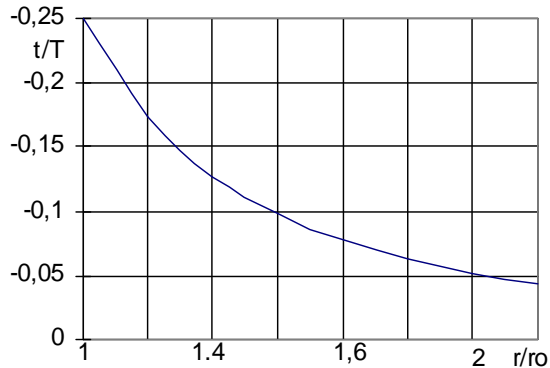

Fig.4. Relative energy $\frac{\hat{t}_{c0}^0}{\overline{T}_0^0}$ as a function of $\frac{r}{r_0}$ for cylindrically symmetric system.

Components $g_{ii}$ of metrical tensor can be calculated from (22), (23), (32), (37), (38) by $M_z = M_{z0}\cdot u_c$.

**8. The conclusion.** Thus, the suggested decision of a problem of gravitational field energy conservation give correct results for important special cases of symmetric static systems. These decisions have simple sense and does not contain contradictions